\documentclass[sigconf,bookmarks=false]{acmart}
\pdfoutput=1
\usepackage{etex}

\copyrightyear{2020} 
\acmYear{2020} 
\setcopyright{rightsretained} 
\acmConference[MSR '20]{17th International Conference on Mining Software Repositories}{October 5--6, 2020}{Seoul, Republic of Korea}
\acmBooktitle{17th International Conference on Mining Software Repositories (MSR '20), October 5--6, 2020, Seoul, Republic of Korea}
\acmDOI{10.1145/3379597.3387460}
\acmISBN{978-1-4503-7517-7/20/05}

\usepackage[T1]{fontenc}
\usepackage{url}
\usepackage{hyperref}
\usepackage[keeplastbox]{flushend}
\usepackage{xspace}
\usepackage{caption}
\usepackage[autolanguage]{numprint}
\usepackage[inline]{enumitem}
\usepackage{ifthen}
\usepackage{balance}

\usepackage{graphicx}
\usepackage{subcaption}

\usepackage{booktabs}
\usepackage{multirow}
\usepackage{threeparttable}
\usepackage{xcolor}
\usepackage{array,makecell}
\usepackage{diagbox}
\usepackage{tabularx}

\usepackage{tcolorbox,tikz}
\usepackage{pgfplots}
\usepackage{pgfplotstable}
\usepackage{filecontents}
\usepgfplotslibrary{statistics}
\usetikzlibrary{calc,patterns,fpu} 
\pgfplotsset{compat=1.14} 
\usepackage{silence}
\pgfplotsset{
    /pgf/declare function={
        Floor(\x) = round(\x-0.49);
    },
    show sum on top/.style={
        /pgfplots/scatter/@post marker code/.append code={%
            \path let \p1=($(normalized axis cs:%
                        \pgfkeysvalueof{/data point/x},%
                        \pgfkeysvalueof{/data point/y})%
                        -(normalized axis cs:\pgfkeysvalueof{/data point/x},0)$)
            in node[
                at={(normalized axis cs:%
                        \pgfkeysvalueof{/data point/x},%
                        \pgfkeysvalueof{/data point/y})%
                },
                anchor={-90*sign(\y1)},yshift={sign(\y1)*2pt}
            ]
            {\pgfmathprintnumber{\pgfkeysvalueof{/data point/y}}};
        },
    }
}

\newcommand*{\boxplott}[7]{%
  \addplot+[
    line width=.2mm, fill, draw=black,
    boxplot prepared={
      lower quartile={#4},
      median={#3},
      upper quartile={#5},
    },
  ] coordinates{};
}

\definecolor{blau_1a}{RGB}{93,133,195}
\definecolor{blau_2a}{RGB}{0,156,218}
\definecolor{gruen_3a}{RGB}{80,182,149}
\definecolor{gruen_4a}{RGB}{175,204,80}
\definecolor{gruen_5a}{RGB}{221,223,72}
\definecolor{orange_6a}{RGB}{255,224,92}
\definecolor{orange_7a}{RGB}{248,186,60}
\definecolor{rot_8a}{RGB}{238,122,52}
\definecolor{rot_9a}{RGB}{233,80,62}
\definecolor{lila_10a}{RGB}{201,48,142}
\definecolor{lila_11a}{RGB}{128,69,151}

\definecolor{blau_1b}{RGB}{0,90,169}
\definecolor{blau_2b}{RGB}{0,131,204}
\definecolor{gruen_3b}{RGB}{0,157,129}
\definecolor{gruen_4b}{RGB}{153,192,0}
\definecolor{gruen_5b}{RGB}{201,212,0}
\definecolor{orange_6b}{RGB}{253,202,0}
\definecolor{orange_7b}{RGB}{245,163,0}
\definecolor{rot_8b}{RGB}{236,101,0}
\definecolor{rot_9b}{RGB}{230,0,26}
\definecolor{lila_10b}{RGB}{166,0,132}
\definecolor{lila_11b}{RGB}{114,16,133}

\newboolean{showcomments}
\setboolean{showcomments}{true}

\ifthenelse{\boolean{showcomments}}
  {\newcommand{\nb}[3]{
  {\color{#2}\small\fbox{\bfseries\sffamily\scriptsize#1}}
  {\color{#2}\sffamily\small$\triangleright~$\textit{\small #3}$~\triangleleft$\GenericWarning{}{LaTeX Warning: #1: #3}}
  }
  \newcommand{\todo}[1]{{\color{red}{TODO: #1}}\GenericWarning{}{LaTeX Warning: TODO: #1}}
    }
  {\newcommand{\nb}[3]{}
  \newcommand{\todo}[1]{}
  }

\usepackage{fp}
\newcommand{\ShowAbsoluteNumber}[1]{%
\ifnum #1<10%
{\hspace*{0pt}#1}%
\else%
\ifnum #1<100%
{\hspace*{0pt}#1}%
\else%
\ifnum #1<1000%
{\hspace*{0pt}#1}%
\else%
{\numprint{#1}}%
\fi%
\fi%
\fi%
}

\newcommand{\ShowPercentage}[2]{%
\FPeval\percentage{round(#1/#2*100,0)}%
\FPeval\percentageOneDecimal{round(#1/#2*100,1)}%
\ifnum \percentage=0%
{\small(\FPprint{percentageOneDecimal}\%)}%
\else%
\ifnum \percentage<10%
{\small(\FPprint{percentageOneDecimal}\%)}%
\else%
{\small(\FPprint{percentage}\%)}%
\fi%
\fi%
\hspace*{0.3ex}%
}
\newlength\BARSIZE  
\newcommand{\inlinechart}[2]{%
\FPeval{\BLACKBARSIZE}{#1/#2}\textcolor{black!80}{\rule{\BLACKBARSIZE\BARSIZE}{1.6ex}}%
\FPeval{\BLACKBARSIZE}{1 - (#1/#2)}\textcolor{black!10}{\rule{\BLACKBARSIZE\BARSIZE}{1.6ex}}%
}

\newcommand*\ChartSmall[2]{%
\numprint{#1}\hspace*{0.5ex}\ShowPercentage{#1}{#2}%
\inlinechart{#1}{#2}%
}
\newcommand{\travis}[0]{Travis~CI\xspace}
\newcommand{\travislistener}[0]{TravisListener\xspace}

\def\nbHistoryBuilds{75139295}

\def\fromDate{Tue, 08 Oct 2019}
\def\toDate{Thu, 19 Dec 2019}
\def\nbRepositories{171057}
\def\nbBuilds{3286773}
\def\nbJobs{9215866}
\def\nbLogs{3007857}

\def\nbRestartedBuilds{56522}
\def\nbRestartedJobs{125461}
\def\nbRestartedLogs{61783}
\def\nbRestartedRepositories{22345}

\def\nbRegex{93}

\def\nbFailureReasons{19730}

\FPeval{percentRestarted}{round(\nbRestartedBuilds*100/\nbBuilds,2)}
\FPeval{percentRestartedRepo}{round(\nbRestartedRepositories*100/\nbRepositories,2)}
\FPeval{percentSuccessfullRestart}{round(\nbRestartedRepositories*100/\nbRepositories,2)}

\FPeval{percentRestartedCanceled}{round(1737*100/\nbRestartedBuilds,2)}
\FPeval{jobsPerBuild}{round(\nbJobs/\nbBuilds,0)}

\def\restartedPassed{46.77}

\FPeval{percentFlaky}{round(\percentRestarted*(\restartedPassed/100),2)}

\FPeval{avgTTBuildRepo}{round(702922/1272,0)}
\FPeval{avgTTJobRepo}{round(3734303/1272,0)}

\newcommand{\answer}[2]{\vspace{.1cm}{\centering\fbox{\parbox{0.97\columnwidth}{\textbf{Answer to RQ#1}. #2}}}\vspace{.2cm}}

\title{Empirical Study of Restarted and Flaky Builds on \travis}

\author{Thomas Durieux}
\affiliation{%
    \institution{INESC-ID and IST, University of Lisbon, Portugal}
}
\email{thomas@durieux.me}

\author{Claire Le Goues}
\affiliation{%
    \institution{Carnegie Mellon University}
}
\email{clegoues@cs.cmu.edu}

\author{Michael Hilton}
\affiliation{%
    \institution{Carnegie Mellon University}
}
\email{mhilton@cmu.edu}

\author{Rui Abreu}
\affiliation{%
    \institution{INESC-ID and IST, University of Lisbon, Portugal}
}
\email{rui@computer.org}

\begin{document}

\begin{abstract}
Continuous Integration (CI) is a development practice where developers frequently integrate code into a common codebase.
After the code is integrated, the CI server runs a test suite and other tools to produce a set of reports (e.g., the output of linters and tests).
If the result of a CI test run is unexpected, developers have the option to manually restart the build, re-running the same test suite on the same code; this can reveal build flakiness, if the restarted build outcome differs from the original build.

In this study, we analyze restarted builds, flaky builds, and their impact on the development workflow.
We observe that developers restart at least \percentRestarted\% of builds, amounting to \numprint{\nbRestartedBuilds}\xspace restarted builds in our \travis dataset.
We observe that more mature and more complex projects are more likely to include restarted builds.
The restarted builds are mostly builds that are initially failing due to a test, network problem, or a \travis limitations such as execution timeout.
Finally, we observe that restarted builds have an impact on development workflow.
Indeed, in 54.42\% of the restarted builds, the developers analyze and restart a build within an hour of the initial build execution. 
This suggests that developers wait for CI results, interrupting their workflow to address the issue.
Restarted builds also slow down the merging of pull requests by a factor of three, bringing median merging time from 16h to 48h.
\end{abstract}

\maketitle

\section{Introduction}

Software engineers use Continuous Integration (CI) not only to integrate their work into a common branch, but also to constantly ensure the quality of their contributions.
Continuous integration was originally introduced in the twelve Extreme Programming practices \cite{beck2000extreme}, and it 
is now considered to a key software development best practice in industry and Open-Source Software alike.

Continuous integration is an automatic process that typically executes the test suite, may further analyze code quality using static analyzers, and can automatically deploy new software versions.
It is generally triggered for each new software change, or at a regular interval, such as daily.
When a CI build fails, the developer is notified, and they typically debug the failing build.
Once the reason for failure is identified, the developer can address the problem, generally by modifying or revoking the code changes that triggered the CI process in the first place.

However, there are situations in which the CI outcome is unexpected, e.g., the build may fail unexpectedly, or for unanticipated reasons.
In this case, developers may \emph{restart} the build (an option generally provided by CI services) to check whether the unexpected outcome is a result of system flakiness or some problem in the CI environment (like network latency).  
A build whose output changes after a restart is referred to as a \emph{flaky build}, because by definition, the software for which a build has been restarted has not changed (but the CI result has). Although there exists prior work studying flaky tests (one contributor to flaky builds)~\cite{luo2014empirical}, to our knowledge, there is no work that studies the practice of build restarts in a CI context, and its implication for development workflow.

Free software and services, such as Jenkins\footnote{Jenkins website: \url{https://jenkins.io/}, visited \today{}} and \travis\footnote{Travis website: \url{https://travis-ci.org/}, visited \today{}}, simplify the adoption of Continuous Integration by the whole community.
In this work we study \travis, one of the largest and most popular 
Continuous Integration services in use today. 
It is free, and its build execution data  is publicly available.
Based on \travis data, we design a study that aims to gain a better understanding of the build restart process and its implication on development workflow.

We have multiple  goals in this study.  
First, we identify how frequently developers restart their builds, and how often restarts surface build flakiness.
Second, we identify the characteristics of projects that restart builds, to understand whether restarting is a common, general practice (or not).
Third, we analyze the reasons that motivate developers to restart their builds.
Finally, we study the impact of restarted builds on developer workflow.
Based on this study, we discuss potential improvements for continuous integration, and ideas for a fully automatized build restart process that could save significant developer time.

Our main observations are that developers restart at least \percentRestarted\% of \travis builds, which represents \numprint{\nbRestartedBuilds}\xspace restarted builds in our dataset.
We observe that the projects that have a restarted build differ from those that do not: more mature and more complex projects are more likely to restart builds. 
We observe that build restart rates differ by project language.
Builds are most commonly restarted in response to test failures, e.g., flaky tests, as well as \travis limitations such as execution timeout.
Finally, we observe that restarted builds appear to have a major impact on developer workflow:
In 53.42\% of the restarted builds, developers initiate a restart within one hour of failure. 
This suggests that developers choose to wait for CI results, interrupting (or pausing) their work to address the issue.
Restarted builds also slow down the merging of pull requests by a factor of three, bringing the median merging time from 16h to 48h.

To summarize, the contributions of this paper are:
\begin{itemize}
    \item A comprehensive study on restarted builds, with implications for CI system design;
    \item A framework, coined \travislistener, to collect real-time data from \travis;
    \item A dataset of \numprint{\nbBuilds}\xspace builds and \numprint{\nbRestartedBuilds}\xspace restarted builds collected in live;
    \item A dataset of \numprint{\nbHistoryBuilds}\xspace builds that cover the complete build history of  \numprint{\nbRepositories}\xspace repositories.
\end{itemize}

The remainder of this paper is organized as follows.
\autoref{sec:background} presents background on \travis and the idea of restarted builds.
\autoref{sec:design} presents study design, including research questions and data collection and analysis.
\autoref{sec:study} presents results, followed by discussion (\autoref{sec:discussion}).
\autoref{sec:threats} presents threats to the validity.
\autoref{sec:related-works} discusses related work, and
\autoref{sec:conclusions} concludes.

\section{Background}\label{sec:background}

In this section, we briefly describe \travis and introduce the notion of restarted builds as well as the technical implications of a restarted build on \travis. 

\subsection{CI and \travis}\label{sec:travis}

\travis is a company that offers an open-source continuous integration service  tightly integrated with GitHub.
It allows developers to build their projects without maintaining their own infrastructure.
\travis provides a simple interface to configure build tasks that are executed for a set of given events: pull requests, commits, crons, and API calls.
Currently, \travis supports 34 programming languages, including Python, Node.js, Java, C, C++, in three operating systems: Linux, Windows, and Mac OSX.
It also supports technologies like Docker, Android apps, iOS apps, and various databases.
The \travis service is free for open-source projects; a paid version is available for private projects.
It is currently used by more than \numprint{932977} open-source projects and \numprint{600000} users.\footnote{Metrics provided at \url{https://travis-ci.org}, visited \today{}}

\travis interacts with GitHub via a set of {webhooks} that are triggered by GitHub events.
For each event, \travis sets up a new {build} by reading the configuration that developers wrote in the repository (.travis.yml file).
Each build is composed of one or several jobs.
A {job} is the execution of the build in a specific environment, for example, one job runs with Java 8 and one with Java 9, or a job can also be used for specific tasks such as deploying Docker images.

\subsection{Restarted Builds}\label{sec:collect_restarted_builds}

A \emph{restarted} build is one that has been relaunched manually, without a corresponding underlying change in the project.  In general, a developer might restart a build when they observe inconsistent or unexpected behavior that they suspect may be due to build or environment flakiness that could be avoided with a new run. 
In this study, we focus on restarted builds from \travis, because 
 \travis is one of the largest continuous integration services, and its build information is also publicly available. 


\begin{figure}[t]
    \centering
    \includegraphics[width=.40\textwidth]{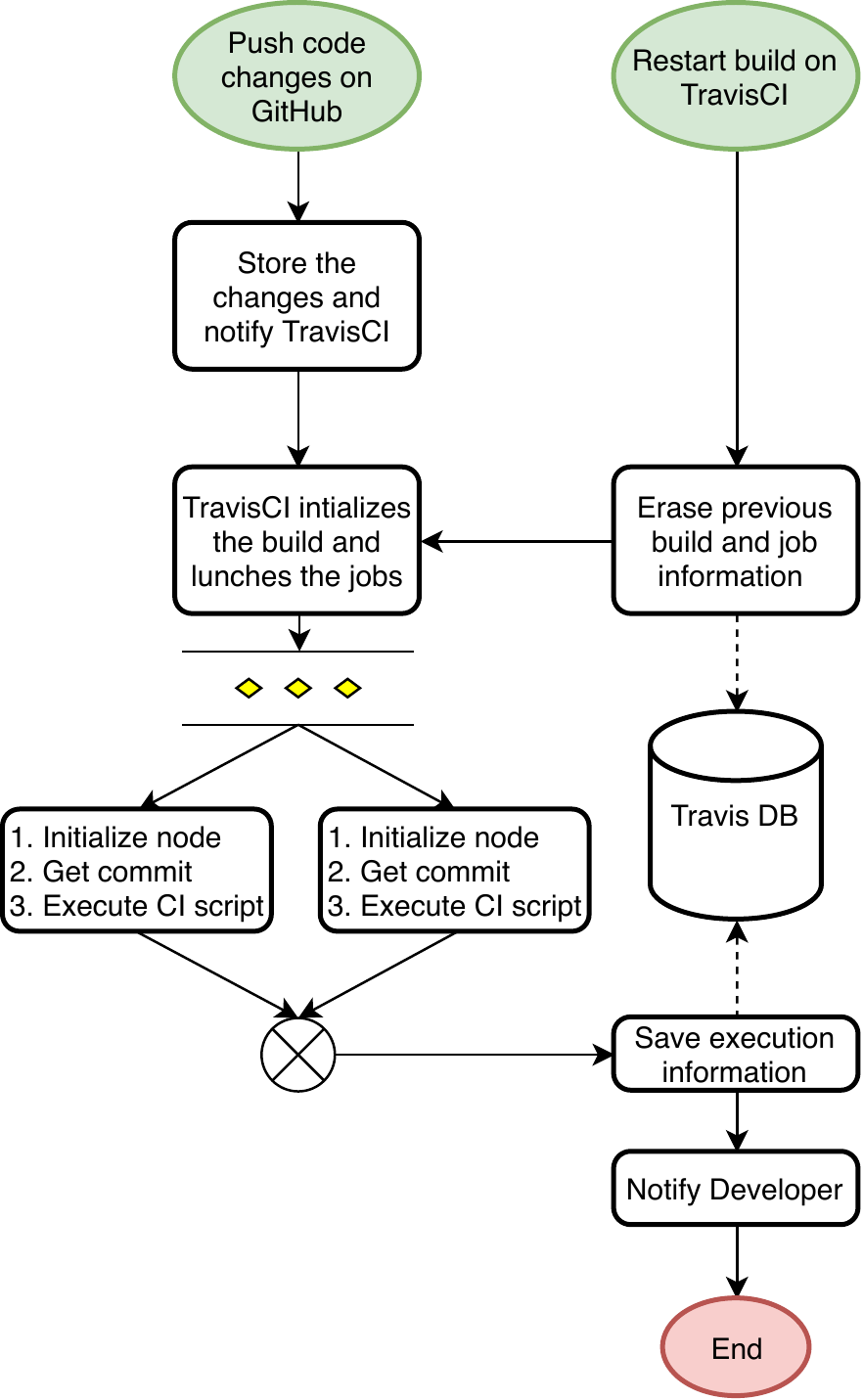}
    \caption{Flowchart of the build restarting process.}
    \label{fig:restart_flowchart}
\end{figure}

In the context of \travis, a restarted build is triggered when a developer clicks on the restart button that the \travis interface provides.
\autoref{fig:restart_flowchart} presents a flowchart of a restarted build on \travis.
The first phase of a restarted build is identical to any build execution on \travis.
First, a developer pushes a code change (commits) to GitHub. 
Second, GitHub processes the commits, and notifies \travis of new code.
Third, \travis analyzes the \travis configuration (\texttt{.travis.yml} file), initializes the different jobs that are configured, and saves the build and job information into its database.
Fourth, \travis executes the configured jobs.
Fifth, \travis saves the execution information in its database (execution time, execution logs).
Finally, \travis notifies the developers that the build execution has ended, and provides build status.

A \emph{restarted} build is triggered directly by the developer, typically by clicking the restart button provided by the \travis web interface (note that \travis also provides an API to restart the build).
Importantly, once a build is restarted, all information saved during the original execution is erased and overwritten.  It is therefore not naively possible to use the \travis API to learn the original status or output of a build that has been restarted. 
Subsequently, execution behavior is identical to a traditional build: \travis configures and executes the jobs, and notifies developers.

The implications of this workflow are first, that developers have to restart a build manually, typically in response to unexpected build behavior. 
Second, the \travis API does not support comparison between a restarted build and its original status. 
We discuss our methodology for performing such comparisons in \autoref{sec:infrastructure}.

\section{Study Design} \label{sec:design}

In this section, we present our study on the restarted builds on \travis and their impact on the developer's workflow.

\subsection{Research Questions}\label{sec:rqs}

During this study, we study four different aspects of restarted builds to answer  the following research questions.
\vspace{-1pt}
\begin{itemize}[leftmargin=24pt]
    \item[\textbf{RQ1}.] How frequently do developers restart builds? How often do restarts identify flaky behavior?  We answer this question by collecting live data from \travis, estimating build restart frequency and identifying how many of those restarted builds have a different status outcome.

    \item[\textbf{RQ2}.] Do the projects that restart builds have the same characteristics as other projects that use \travis?
    We answer this question by studying the characteristics of projects that restarted builds over the period of our live study. We aim to identify potential common characteristics of those projects.
    
    \item[\textbf{RQ3}.] Why do developers restart builds?
    In the third research question, we analyze build logs to the extract reasons for the failures of builds that developers choose to restart.  
    
    \item[\textbf{RQ4}.] What is the impact of flaky builds on development workflow?
    In this final research question, 
    we study when (temporally) developers restart build, as well as the relationship between build restarts and pull requests.  
\end{itemize}

\subsection{\travislistener Infrastructure}\label{sec:infrastructure}

\begin{figure}
    \centering
    \includegraphics[width=0.4\textwidth]{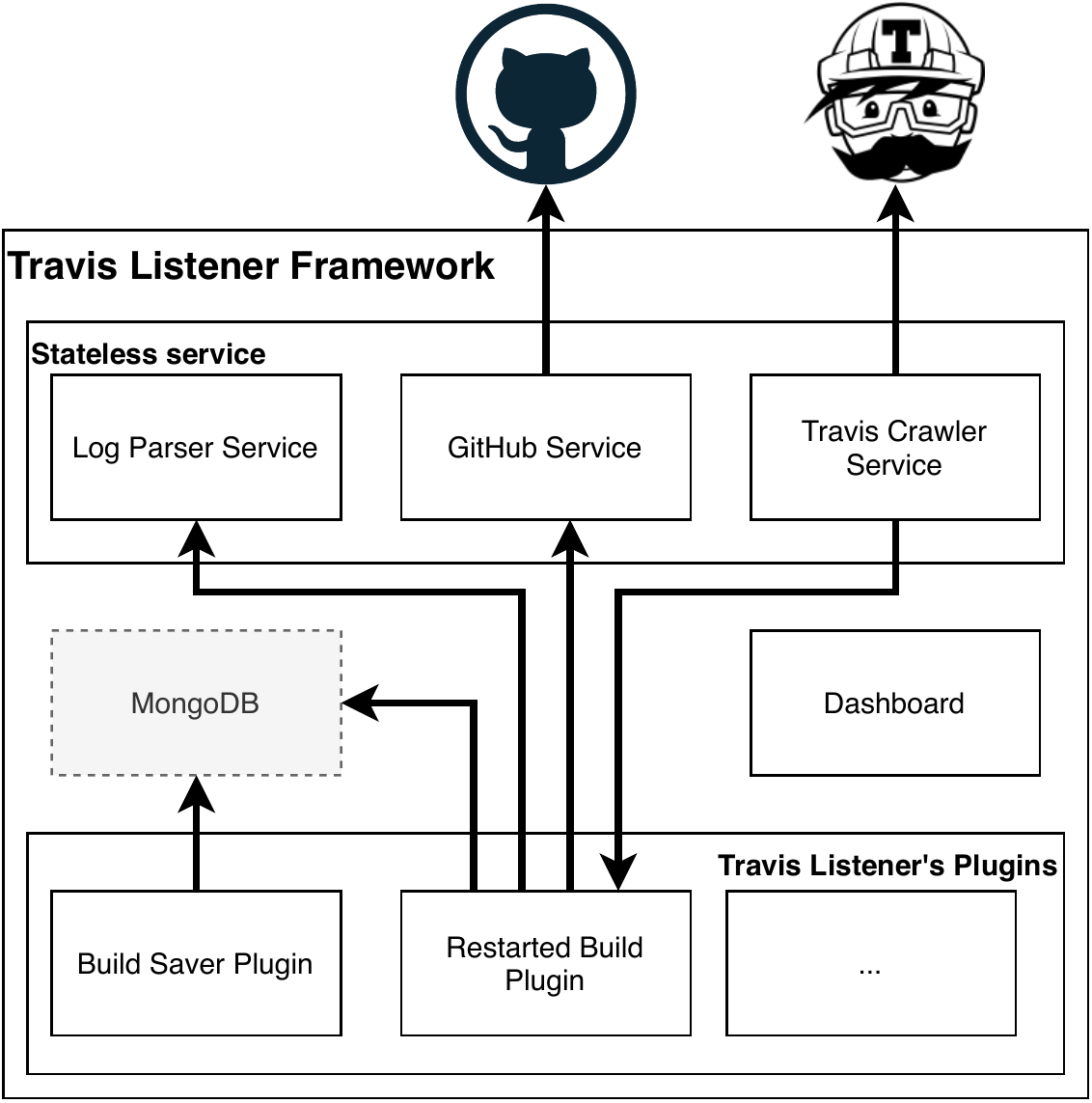}
    \caption{Architecture of our \travislistener framework.}
    \label{fig:travislistener_infrastructure}
\end{figure}

\autoref{fig:travislistener_infrastructure} depicts our infrastructure for collecting the data required to answer our four research questions.
The infrastructure consists of seven modules: three services, two plugins, a dashboard, and a database.
The infrastructure is built on top of Docker compose v2.4.\footnote{Docker Compose documentation \url{https://docs.docker.com/compose/} visited \today{}}
Docker Compose is straightforward to install, scalable, and resilient (given, e.g., its auto-restart capabilities). 
Each module of our infrastructure is thus a docker image integrated into  Docker compose.
The services, plugins, and the dashboard are implemented using JavaScript and Node.js v.10.

The \textbf{Dashboard} provides a web interface to configure and monitor the state of the different modules of the system.

For the \textbf{Database}, we use MongoDB\footnote{MongoDB website: \url{https://mongodb.github.io/} visited \today{}} which integrates well with Node.js and provides data compression by default.
Data compression is a useful feature, since we collect millions of highly compressible log files.

The \textbf{Log Parser Service} is a service that is used to manipulate logs. 
The current version of the service provides the following features: 
1) Log minimization: removes the meaningless content such as progress bar status and log formatting. 
2) Log Diff: produced minimized diffs between two logs by removing all random or time-based content, such as ids or dates. 
3) Data extraction: parses the log to extract failures reasons such as test failures, checkstyle warnings, compilation errors, or timeouts. 
We are currently using \nbRegex\xspace regular expressions to extract failure reasons from logs.

The \textbf{GitHub Service} is a simple middleware component that handles GitHub API's tokens.
It serves to simplify the usage of the GitHub API within \travislistener by centralizing identification and rate limiting.

The \textbf{Travis Crawler Service} extracts the information from \travis. 
Its main purpose is to crawl \travis to detect any new jobs and builds triggered by \travis, live. 
\textbf{Travis Crawler Service} provides a WebSocket service that can be listened to by all \travislistener modules. 
The WebSocket provides live notifications for any new \travis jobs or builds.

The \textbf{Build Saver Plugin} listens to the \textbf{Travis Crawler Service} and saves all information to the database.
We save the following information: \travis's job, \travis's build, commit information (not including the diff), repository information, and user information.
The goal of this plugin is to track all changes, and provide statistics on who is using \travis.

The \textbf{Restarted Build Plugin} collects the information relevant to the present study.  
Its goal is to detect restarted builds on \travis.
As explained in \autoref{sec:collect_restarted_builds}, when a build is restarted by a developer, all the original information is overwritten.  Tracking restarted builds thus requires live collection of build data (in our case, using the \textbf{Build Saver Plugin}).
To detect restarted builds, the \textbf{Restarted Build Plugin} crawls periodically (once a day) the collected builds from the 30 previous days, comparing the build start timestamp provided by the \travis' API to the start time saved by the \textbf{Build Saver Plugin}. 
If the two times differ, the build was restarted.
For each restarted build, we collect the new \travis job information and execution logs.

The modularity and the use of Docker images make the framework highly flexible. 
We develop this live infrastructure with the vision that it will be reused by other developers for other purposes.
It could be extended to monitor other continuous integration services, other Git services, or even to provide different services.
For example, we envision that \travislistener could easily be extended to create a production service that monitors \travis and automatically provides feedback to developers.
The source code of \travislistener is available on GitHub: \url{https://github.com/tdurieux/Travis-Listener}.

\subsection{Data collection overview}\label{sec:data_collection}

\begin{table*}[t]
\minipage[t]{0.32\textwidth}%
    \caption{Overview of the collected data.}
    \label{tab:main_metric}
    \centering
    \begin{tabular}{@{}l r@{}}\toprule
Metric Name & Value \\\midrule
Start Study & \fromDate \\
End Study & \toDate \\
\# Builds & \numprint{\nbBuilds} \\
\# Jobs & \numprint{\nbJobs} \\
\# Logs & \numprint{\nbLogs} \\
\# Repositories & \numprint{\nbRepositories} \\
\midrule
\# Passed builds & \ChartSmall{2306130}{\nbBuilds} \\
\# Failed builds & \ChartSmall{583415}{\nbBuilds} \\
\# Errored builds & \ChartSmall{367963}{\nbBuilds} \\
\# Canceled builds & \ChartSmall{29265}{\nbBuilds} \\
\midrule
Avg. \# builds per repo & \numprint{19} \\
Avg. \# jobs per repo & \numprint{52} \\
\bottomrule
    \end{tabular}
\endminipage\hfill
\minipage[t]{0.32\textwidth}%
    \caption{Overview of the restarted builds.}
    \label{tab:restarted_metric}
    \centering
    \begin{tabular}{@{}l r@{}}\toprule
Metric Name & Value \\\midrule
\# Restarted Builds & \ChartSmall{\nbRestartedBuilds}{\nbBuilds} \\
\# Restarted Jobs & \ChartSmall{\nbRestartedJobs}{\nbJobs} \\
\# Restarted Logs & \ChartSmall{\nbRestartedLogs}{\nbLogs} \\
\# Restarted Repositories & \ChartSmall{\nbRestartedRepositories}{\nbRepositories} \\
\midrule
\# Restart Passed builds & \ChartSmall{9707}{\nbRestartedBuilds} \\
\# Restart Failed builds & \ChartSmall{27006}{\nbRestartedBuilds} \\
\# Restart Errored builds & \ChartSmall{17621}{\nbRestartedBuilds} \\
\# Restart Canceled builds & \ChartSmall{2186}{\nbRestartedBuilds} \\
\bottomrule
    \end{tabular}
\endminipage\hfill
\minipage[t]{0.32\textwidth}%
    \caption{Overview of TravisTorrent metrics.}
    \label{tab:travistorrent}
    \centering
    \begin{tabular}{@{}l r@{}}\toprule
Metric Name & Value \\\midrule
Start Study & 2011-08-29 \\
End Study & 2016-08-31 \\
\# Builds & \numprint{702922} \\
\# Jobs & \numprint{3734303} \\
\# Repositories & \numprint{1272} \\
\# Languages & \numprint{3} \\
\midrule
\# Passed builds & \ChartSmall{524017}{702922} \\
\# Failed builds & \ChartSmall{124821}{702922} \\
\# Errored builds & \ChartSmall{55907}{702922} \\
\# Canceled builds & \ChartSmall{2394}{702922} \\
\midrule
Avg. \# builds per repo & \numprint{\avgTTBuildRepo} \\
Avg. \# jobs per repo & \numprint{\avgTTJobRepo} \\
\bottomrule
\end{tabular}
\endminipage
\end{table*}

Data collection is composed of two main steps. The first step uses \travislistener (see \autoref{sec:infrastructure}) to collect restarted builds from \travis. 
The second step collects additional information from \travis and GitHub to characterize the projects.

To recap, the first step is the collection of the builds and jobs in real-time from \travis between \fromDate\xspace and \toDate.
We save each new build and job from \travis in our MongoDB database.
When a failing job is detected, we download its execution log.
The log is cleaned to reduce its size, and then stored.
Every hour, we query \travis to request the build information from the last 30 days. 
We compare the starting date from the freshly collected data with the data that we previously stored in our database.
If the build information is different between the two collections (primarily the starting timestamp), it means the build has been restarted.
For each restarted build, we collect the new job information and  execution logs.
We then extract the failure reason using regular expressions, and save everything in the database.

In the second step, we collected data from \travis and GitHub to study the different aspects of the repositories.
We collected the complete \travis build history of all repositories that have triggered a build during our study period.
The complete history contains \numprint{74981298} builds, amounting to 32G.
We also collect GitHub repository information, which contains the programming language, the number of stars, the creation date, and the repository size.
We collect pull requests comments for all pull requests that have at least one build collected during our study.
Finally, we collect the commit history for all the repositories that have at least one restarted build.

\autoref{tab:main_metric} presents primary statistics of the collected data.
In total, we collected \numprint{\nbBuilds}\xspace builds, \numprint{\nbJobs}\xspace jobs, \jobsPerBuild\xspace jobs per build on average, from  \numprint{\nbRepositories}\xspace repositories. 
During the collection procedure, we observed that 70\% of the builds pass, 18\% fail, 11\% error, and 0.9\% are canceled.

\autoref{tab:restarted_metric} presents the main statistics of the restarted builds that we identify during our study period.
In total, we identify \numprint{\nbRestartedBuilds}\xspace restarted builds, \numprint{\nbRestartedJobs}\xspace restarted jobs on \numprint{\nbRestartedRepositories}\xspace repositories.
Among the restarted builds, 17\% of the builds were passing, 48\% were failing, 31\% errored, and 3.9\% were canceled.

\autoref{tab:travistorrent} presents the same metrics, but from TravisTorrent dataset, which we provide for context and comparison.  
Our dataset differs meaningfully from TravisTorrent. 
First, because of the live study period, we focus on a short, recent time period; TravisTorrent's dataset spans a wider, older timeframe. 
Second, TravisTorrent contains the builds from \numprint{1272} repositories, from three different programming languages;  we have builds from \numprint{\nbRepositories}\xspace repositories and all 34 supported languages.
The two datasets do not have the same goal: TravisTorrent focuses on a few projects but covers a large portion of the build history of those projects, while our dataset focuses on the diversity of projects and languages. 

We create this new dataset for two reasons.  First, and most important, TravisTorrent does not contain information about restarted builds, and the information cannot be collected post-facto (as described above).  Second, our dataset provides a larger diversity of projects and languages than other previous datasets. 

To summarize, we live-collected \numprint{\nbBuilds}\xspace builds, \numprint{\nbJobs}\xspace jobs and \numprint{\nbLogs}\xspace logs.
We detected \numprint{\nbRestartedBuilds}\xspace restarted builds. 
We download the complete \travis history of \numprint{\nbRepositories}\xspace projects, representing \numprint{\nbHistoryBuilds}\xspace builds.
The complete uncompressed size of the dataset is approximately 500Gb.
The collected data is available on Zenodo with the following DOI: \href{https://doi.org/10.5281/zenodo.3601137}{10.5281/zenodo.3601137}. 

\section{Study on Restarted Builds}
\label{sec:study}

In this section, we answer our four research questions.

\subsection{RQ1. Restarted Builds}

In this first research question, we investigate how many builds are restarted by developers. 
The goal of this first research question is to determine if developers are actually restarting builds; if so, how often; and, whether the restarted builds change in outcome post-restart.  

\autoref{tab:restarted_metric} presents the main pieces of evidence to answer this first research question. 
We identified \numprint{\nbRestartedBuilds}\xspace restarted builds, representing \percentRestarted\% of all the builds that we observed during our study.
This indicates that developers are indeed restarting builds. Note that our observation underestimates reality, because we miss some of the restarted builds.
Indeed, we are not able to detect that builds that are restarted too quickly because we need to collect the original state of the build and the restarting state of the build.
Our \travis crawler is blind to builds that are restarted a few seconds after the original.  We therefore do not consider such builds.
We are also missing builds that are restarted after 30 days, primarily to to reduce the number of requests made on \travis.
We are surprised to observe that so many builds are restarted, primarily because doing so requires manual labor from a developer with sufficient project privileges; intuitively, we expected that a privileged, manual procedure would appear only marginally throughout a fully automatic process like CI.

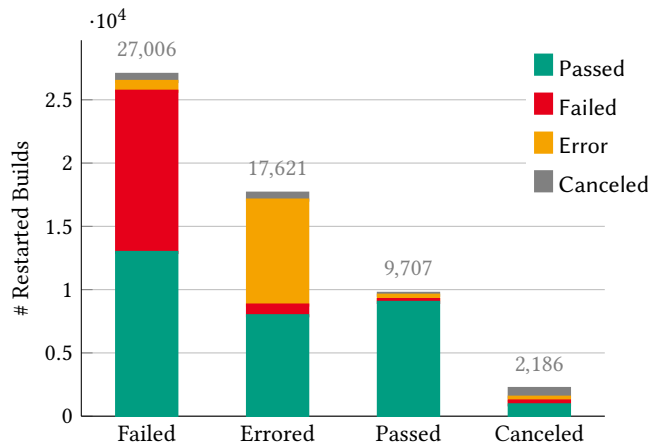
\begin{figure}[t] 
	\centering
	\begin{tikzpicture} 
		\begin{axis}[
			ybar stacked,
			xmin = 0.5,
			xmax = 4.5,
			ymin = 0,
			axis x line* = bottom,
			axis y line* = left,
			ylabel= \# Restarted Builds,
			width= 0.48\textwidth,
			height = 0.37\textwidth,
			ymajorgrids = true,
			bar width = 8mm,
			xticklabels = \empty,
			extra x ticks = {1,2,3,4},
			extra x tick labels = {Failed, Errored, Passed, Canceled, },
			nodes near coords={},
            legend style={ 
               row sep=3pt, 
               draw=none, 
               anchor=north,
               cells={anchor=west,font=\sffamily}},
			]
			
            \addplot+[mark=none, gruen_3b, very thick] coordinates {(1, 12936) (2, 7937) (3, 8991) (4, 898) };
            \addplot+[mark=none, rot_9b, very thick] coordinates {(1, 12739) (2, 843) (3, 223) (4, 302) };
            \addplot+[mark=none, orange_7b, very thick] coordinates {(1, 774) (2, 8295) (3, 347) (4, 300) };
            \addplot+[mark=none, gray, very thick,show sum on top] coordinates {(1, 557) (2, 546) (3, 146) (4, 686) };

			\addlegendentry{Passed}
            \addlegendentry{Failed}
            \addlegendentry{Error}
            \addlegendentry{Canceled}
		\end{axis} 
	\end{tikzpicture}
	\caption{Evolution of the build status between original builds and restarted builds.}
	\label{fig:restarted_state}
\end{figure}

Our overall hypothesis is that developers restart a build because they expect the build behavior to change.  
\autoref{fig:restarted_state} corroborates our hypothesis.
This figure shows how the states of the builds change between the original build and the restarted build.
The horizontal axis presents the original state; the vertical axis presents the numbers of restarted builds; and the stacked colors represent the state of the restarted builds.
We observe that \restartedPassed\% of the failing/errored builds pass post-restart.
This high percentage of state changes between the original and the restarted builds surprised us, because it indicates that builds are flaky; it also indicates that developers are reasonably adept at identifying such flaky builds.
In our study, we observed that \percentFlaky\% (\restartedPassed\% of \percentRestarted\% of the restarted builds) of all \travis builds in our dataset are flaky (and recall that this is a lower bound).
We study the cause of flakiness in greater depth in the third research question (see \autoref{sec:rq3}).

\begin{figure}[t] 
	\centering
	\begin{tikzpicture} 
		\begin{axis}[
			ybar,
			xmin = 0.5,
			xmax = 7.5,
			ymin = 0,
            bar width = 6.5mm,
			axis x line* = bottom,
			axis y line* = left,
			ylabel= \# Repositories,
			xlabel= \# Restarted Builds,
			width= 0.48\textwidth,
			height = 0.32\textwidth,
			point meta={y*100/\nbRestartedRepositories},
			visualization depends on=rawy \as \myy,
            nodes near coords={\pgfmathprintnumber\pgfplotspointmeta\%},
			every node near coord/.style={
			    color = gray,
			},
			ymajorgrids = true,
			xticklabels = \empty,
            extra x ticks = {1,2,3,4,5,6,7},
			extra x tick labels = {1, 2-5, 5-10, 10-15, 15-20, 20-50, > 50},
			]
			\addplot+[mark=none, blau_2b, very thick] coordinates {(1, 13305) (2, 7500) (3, 1130) (4, 285) (5, 124) (6, 168) (7, 42) };
		\end{axis} 
	\end{tikzpicture}
	\caption{Distribution of the number of repositories versus the number of time the repositories restart a build.}
	\label{fig:nb_restarted_project}
\end{figure}
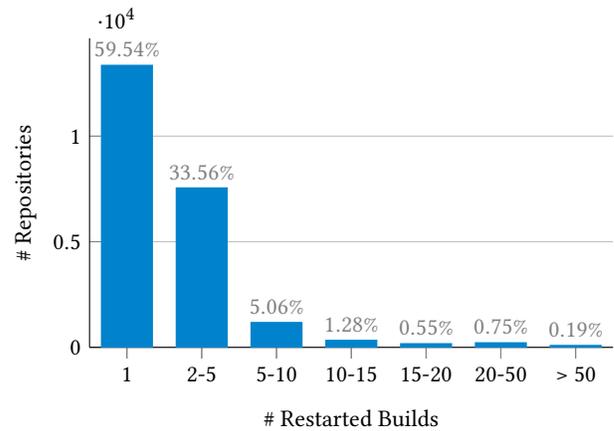

In total, \numprint{\nbRepositories}\xspace repositories have at least one restarted build, which represents \percentRestartedRepo\% of all the repositories that triggered a build during our study period.
Note that most repositories that include restarts only restarted a few builds. 
This is confirmed by \autoref{fig:nb_restarted_project}, which presents how frequently projects restart builds. 
This chart shows that 59.54\% of the repositories restart only one build, and 33.56\% restart between 2-5 builds.
Restarting a build is, overall, a rare operation in the development workflow.
However, some projects restart builds far more frequently.  For example, the project getsentry/sentry\footnote{\url{https://github.com/getsentry/sentry}} restarted 231 builds (out of \numprint{1613}) during our study period. 
This represents a restart rate of 14\%, much higher than the average that we observed.
One possible reason for this this high number of restarted builds is a high number of flaky tests; indeed, we found 33 pull requests to this project that mention flaky tests. 

\answer{1}{
\textbf{How frequently do developers restart builds? How often do restarts identify flaky behavior?}
We observe that developers indeed restart builds.
In our study, we observe that \percentRestarted\% of the builds are restarted.
However, restarting a build is not a common task, and most of the projects only restart one to five builds during the studied time period. 
Interestingly, \restartedPassed\% of the restarted builds change their failing state to a passing state after the restart.
This suggests that the restarted builds suffer from flaky behavior.}

\subsection{RQ2. Projects that restart builds}

In this second research question, we analyze the general characteristics of the projects that restart builds.
The goal is to identify patterns that differentiate the restarted builds from the non-restarted ones.
We analyze: 
\begin{enumerate*}
    \item the programming languages, 
    \item number of builds triggered by the repositories, 
    \item execution time, and finally
    \item popularity of the repository (number of stars).
\end{enumerate*}

\begin{figure}[t] 
	\centering
	\begin{tikzpicture} 
		\pgfplotsset{
      scale only axis,
  }

  \begin{axis}[
  	ybar,
    axis y line*=left,
			xmin = 0.5,
			xmax = 10.5,
			ymin = 0,
			axis x line* = bottom,
			axis y line* = left,
			ylabel= Proportion of restarted builds,
			width= 0.40\textwidth,
			height = 0.28\textwidth,
			ymajorgrids = true,
			bar width = 4mm,
			xticklabels = \empty,
			extra x ticks = {1,2,3,4,5,6,7,8,9,10},
			extra x tick labels = {Python, JavaScript, Java, C++, Go, PHP, TypeScript, C, Ruby, Shell,},
			extra x tick style={
              tick label style={rotate=90}
            },
			visualization depends on=rawy \as \myy,
            nodes near coords={\pgfmathprintnumber\pgfplotspointmeta\%},
			every node near coord/.style={
			    color = gray,
			    rotate=90,
			    anchor=west
			},
            legend style={ 
               row sep=3pt, 
               draw=none, 
               cells={anchor=west,font=\sffamily}},
  ]

            \addplot+[mark=none, blau_2b, very thick] coordinates {(1, 1.720163898597385) (2, 1.0191875038646125) (3, 1.961235317569847) (4, 2.8176830713554826) (5, 2.067070995362112) (6, 1.5119672387082828) (7, 1.0483476720330591) (8, 2.105181537984376) (9, 1.4769427544247788) (10, 1.2725200319831238) };

		\end{axis} 
	\end{tikzpicture}
	\caption{The restarted rate per language.}
	\label{fig:rate_languages}
\end{figure}
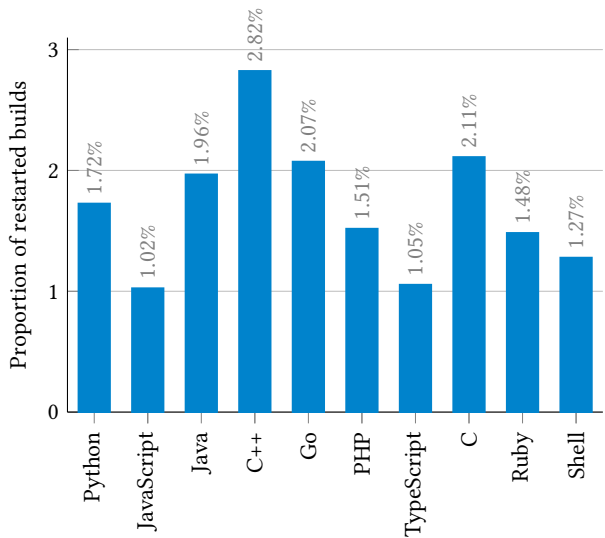
\autoref{fig:rate_languages} presents the restarting rate per language for the ten most frequent languages in \travis.
JavaScript and TypeScript stand out.
Only 1.02\% of the JavaScript builds, and 1.05\% of the TypeScript builds are restarted, whereas the other languages have build restart rates between 1.27 and 2.82\%.
To seek to explain this difference, we analyze the distribution of build state between the languages (see \autoref{fig:state_languages}), the number of builds for each languages (see \autoref{fig:nbBuilds}) and the build execution times \autoref{fig:duration}).
This analysis is based on the complete \travis build history of \numprint{\nbRepositories}\xspace projects, which represents a total of \numprint{\nbHistoryBuilds}\xspace builds.

\autoref{fig:nbBuilds} presents the success rate for the top 10 programming languages. 
The boxes with vertical lines present the success rate of the projects that did \emph{not} restart builds in the studied time period.
The boxes without vertical lines are the success rate for the projects that have at least one restarted build.
A higher success rate for JavaScript and TypeScript would explain the lower number of restarted builds.
However, as this figure attests, we did not observe any major differences between the success rate of JavaScript and TypeScript and the other languages.

\autoref{fig:nbBuilds} presents a box chart of the median number of \travis builds per project for each language.
The gray box presents the value for the projects that do not have a restarted build, and the blue bar presents the value for the projects that have at least one restarted build. 
We also did not identify a significant difference between JavaScript/TypeScript and other languages.

Finally, \autoref{fig:duration} presents box plots of the median build duration for each language. 
The gray box presents the value for the projects that do not have a restarted build, and the blue bar presents the value for the projects that have at least one restarted build. 
Here, we do observe a difference between JavaScript and TypeScript and other languages: 
JavaScript and TypeScript builds are substantially faster than those in other programming languages.
A shorter build execution time typically suggests that the builds are less complex than those for other languages and, therefore, likely less prone to producing unexpected behavior requiring restarts.

\begin{figure*}
\minipage[t]{0.325\textwidth}%
\begin{tikzpicture} 
\begin{axis}[
    xbar stacked,
	ymin = 0.65,
	ymax = 10.65,
	xmin = 0,
 	xmax = 100,
	axis x line* = bottom,
	axis y line* = left,
	xlabel= \% Build Status,
	width= \textwidth,
	height = 1.3\textwidth,
	xmajorgrids = true,
	bar width = 1.3mm,
	yticklabels = \empty,
	extra y ticks = {1.1,2.1,3.1,4.1,5.1,6.1,7.1,8.1,9.1,10.1},
	extra y tick labels = {Python, JavaScript, Java, C++, Go, PHP, TypeScript, C, Ruby, Shell,},
	legend columns=4,
    legend style={
        row sep=3pt,
        draw=none,
        at={(0.33,-0.14)},
        anchor=north,
        cells={anchor=west,font=\sffamily}
    },
]

\addplot+[mark=none, draw=none, gruen_3b, very thick] coordinates {(61, 1) (66, 2) (64, 3) (59, 4) (70, 5) (65, 6) (68, 7) (64, 8) (65, 9) (62, 10) };
\addplot+[mark=none, draw=none, rot_9b, very thick] coordinates {(25, 1) (18, 2) (18, 3) (20, 4) (18, 5) (20, 6) (18, 7) (19, 8) (21, 9) (21, 10) };
\addplot+[mark=none, draw=none, orange_7b, very thick] coordinates {(10, 1) (12, 2) (15, 3) (13, 4) (9, 5) (11, 6) (10, 7) (11, 8) (11, 9) (10, 10) };
\addplot+[mark=none, draw=none, gray, very thick] coordinates {(5, 1) (4, 2) (4, 3) (8, 4) (3, 5) (4, 6) (4, 7) (7, 8) (4, 9) (7, 10) };
    
\addlegendentry{Passed}
\addlegendentry{Failed}
\addlegendentry{Error}
\addlegendentry{Canceled}
		
	\end{axis} 
    \begin{axis}[
    	xbar stacked,
        axis x line*=left,
        axis y line=none,
		ymin = 0.35,
		ymax = 10.35,
        xmin = 0,
 		xmax = 100,
        bar width = 1.3mm,
		width= \textwidth,
		height = 1.3\textwidth,
        nodes near coords={},
    ]   
\addplot+[mark=none, draw=none, gruen_3b, very thick, postaction={
    pattern=vertical lines
}] coordinates {(66, 1) (74, 2) (69, 3) (63, 4) (74, 5) (69, 6) (74, 7) (69, 8) (68, 9) (71, 10) };
\addplot+[mark=none, draw=none, rot_9b, very thick, postaction={
    pattern=vertical lines
}] coordinates {(22, 1) (15, 2) (14, 3) (21, 4) (14, 5) (17, 6) (15, 7) (18, 8) (17, 9) (17, 10) };
\addplot+[mark=none, draw=none, orange_7b, very thick, postaction={
    pattern=vertical lines
}] coordinates {(9, 1) (9, 2) (15, 3) (12, 4) (9, 5) (12, 6) (9, 7) (9, 8) (13, 9) (8, 10) };
\addplot+[mark=none, draw=none, gray, very thick, postaction={
    pattern=vertical lines
}] coordinates {(3, 1) (2, 2) (2, 3) (4, 4) (2, 5) (2, 6) (2, 7) (4, 8) (2, 9) (4, 10) };

\end{axis} 
\end{tikzpicture}
\caption{Comparison of the distribution of build status between the projects with restarted builds (plain box) vs. the other projects (boxes with vertical lines).}
\label{fig:state_languages}
\endminipage\hfill
\minipage[t]{0.325\textwidth}%
\begin{tikzpicture}
\begin{axis}[
    boxplot/draw direction=x,
    xmin=0,
    ymin=0,
    ymax=10,
    width=\textwidth,
    height = 1.3\textwidth,
    axis x line* = bottom,
    axis y line* = left,
    xlabel= \# Builds,
    xmajorgrids = true,
    xtick style={draw=none}, 
    ytick style={draw=none}, 
    cycle list={{blau_2b},{gray}},
    boxplot={
            draw position={1/3 + Floor(\plotnumofactualtype/2) + 1/3*mod(\plotnumofactualtype,2)},
            box extend=0.3
    },
    ytick={0,...,10},
    y tick label as interval,
    yticklabels={Python, JavaScript, Java, C++, Go, PHP, TypeScript, C, Ruby, Shell,},
    legend style={
        row sep=3pt,
        draw=none,
        legend columns=-1,
        at={(0.5,-0.14)},
        anchor=north,
        cells={anchor=west,font=\sffamily}
    },
    legend image code/.code={
        \draw[#1, draw=none] (0cm,-0.1cm) rectangle (0.2cm,0.1cm);
    },
]

\addlegendentry{Restarted\;};
\addlegendentry{Non-Restarted\;};

\boxplott{1}{1391.8546790162886}{274}{49}{1174}{1}{66674}
\boxplott{1}{356.7085697474047}{99}{25}{324}{1}{170749}

\boxplott{2}{1117.7590822179732}{149}{20}{824}{1}{67649}
\boxplott{2}{330.93315402979664}{99}{25}{324}{1}{187699}

\boxplott{3}{1055.0542420027816}{149}{25}{824}{1}{45340}
\boxplott{3}{348.4410850760057}{99}{25}{324}{1}{39174}

\boxplott{4}{1687.0428462127009}{474}{99}{1486.5}{1}{47574}
\boxplott{4}{426.903443502336}{124}{25}{374}{1}{72324}

\boxplott{5}{1419.2648125755743}{299}{49}{1349}{1}{43099}
\boxplott{5}{284.48200132100396}{74}{25}{274}{1}{37799}

\boxplott{6}{1652.9539406345957}{224}{49}{924}{1}{106574}
\boxplott{6}{326.9698254609999}{99}{25}{299}{1}{70474}

\boxplott{7}{1314.0686390532544}{199}{25}{949}{1}{98899}
\boxplott{7}{340.8888888888889}{99}{25}{374}{1}{17124}

\boxplott{8}{1836.771212121212}{462}{99}{1605.25}{1}{43549}
\boxplott{8}{463.31150910158544}{149}{25}{449}{1}{121399}

\boxplott{9}{2088.5400696864112}{499}{124}{1692.75}{1}{77424}
\boxplott{9}{450.9194444444444}{124}{25}{374}{1}{128174}

\boxplott{10}{1295.1986644407345}{174}{49}{599}{1}{281874}
\boxplott{10}{386.2625382623028}{99}{25}{324}{1}{169899}

\end{axis}
\end{tikzpicture}
\caption{Box plot of the number of builds for the projects that have restarted builds (blue box) vs. the projects that do not have (gray box). It shows that the projects with restated builds have more builds that the other projects.}
\label{fig:nbBuilds}
\endminipage\hfill
\minipage[t]{0.325\textwidth}%
\begin{tikzpicture}
\begin{axis}[
    boxplot/draw direction=x,
    xmin=0,
    ymin=0,
    ymax=10,
    width=\textwidth,
    height = 1.3\textwidth,
    axis x line* = bottom,
    axis y line* = left,
    xlabel= Build Duration in sec,
    xmajorgrids = true,
    xtick style={draw=none}, 
    ytick style={draw=none}, 
    cycle list={{blau_2b},{gray}},
    boxplot={
        draw position={1/3 + Floor(\plotnumofactualtype/2) + 1/3*mod(\plotnumofactualtype,2)},
        box extend=0.3
    },
    ytick={0,...,10},
    y tick label as interval,
    yticklabels={Python, JavaScript, Java, C++, Go, PHP, TypeScript, C, Ruby, Shell,},
    legend style={
        row sep=3pt,
        draw=none,
        legend columns=-1,
        at={(0.5,-0.14)},
        anchor=north,
        cells={anchor=west,font=\sffamily}
    },
    legend image code/.code={
        \draw[#1, draw=none] (0cm,-0.1cm) rectangle (0.2cm,0.1cm);
    }
]

    
\addlegendentry{Restarted\;};
\addlegendentry{Non-Restarted\;};

\boxplott{1}{424.20089428297666}{207}{78}{502.5}{13}{9694}
\boxplott{1}{221.46019670477472}{97}{55}{209}{2}{22339}

\boxplott{2}{264.99393746011486}{134}{73.625}{296.875}{9}{11163}
\boxplott{2}{140.55792019347038}{84}{57}{139}{5}{10422}

\boxplott{3}{440.99420222634507}{171}{83}{451.75}{13}{8676}
\boxplott{3}{264.5919438844086}{118}{69}{234}{6}{12585}

\boxplott{4}{1184.4912012241775}{829}{317.5}{1681.75}{12}{21222}
\boxplott{4}{657.2554517133956}{317.75}{124.625}{780}{12}{23646}

\boxplott{5}{371.5072727272727}{220}{110}{431}{15}{5670}
\boxplott{5}{195.73150594451783}{128}{70}{238}{9}{5405}

\boxplott{6}{339.32497441146364}{186}{99}{389.5}{13}{5964}
\boxplott{6}{211.1926718611303}{121.5}{76.5}{205.25}{6}{14447.5}

\boxplott{7}{295.86272189349114}{179}{98.5}{336}{14}{2908}
\boxplott{7}{154.86558930425753}{102}{67}{168}{12}{3654}

\boxplott{8}{1107.2015151515152}{629.75}{263.375}{1265}{15}{16280}
\boxplott{8}{581.2196124486201}{199.75}{75}{549.375}{6}{16982}

\boxplott{9}{447.56881533101046}{279}{130}{511.75}{17}{4500}
\boxplott{9}{244.98880654624054}{139}{83}{276}{15}{17697}

\boxplott{10}{702.329716193656}{362.5}{120.5}{956.75}{15}{5752}
\boxplott{10}{391.88742345737165}{140}{55}{424.375}{10}{9599}

\end{axis}
\end{tikzpicture}
\caption{Box plots of the execution time of projects that have restarted builds (blue box) vs. the projects that do not have (gray box). It shows that the execution time of projects that have restarted builds are much higher.} 
\label{fig:duration}
\endminipage
\end{figure*}


\begin{table}[t]
\centering
\caption{Number of stars of \numprint{21775} projects that have restarted builds compared to the \numprint{153666} that do not have a restarted build. The number of projects differs from \autoref{tab:main_metric} and \autoref{tab:restarted_metric} because we lack data on the repositories that are no longer accessible since initial data collection.}
\label{tab:stars}
\begin{tabular}{@{}l r r@{}}
\toprule
Metric           & Restarted & Non-Restarted \\
\midrule
\# Stars & \numprint{8256003} & \numprint{20837112} \\
Average & \numprint{379} & \numprint{136} \\
Median & \numprint{3} & \numprint{1} \\
\bottomrule
\end{tabular}
\end{table}

Turning to project popularity:
\autoref{tab:stars} presents the number of stars for the projects with a restarted build compared to the other projects.
We observe that the projects with restarted builds are more popular than the others. 
They have, on average, 379 stars, with a median of 3 stars. 
The other projects only have an average of 136 stars and a median of 1.

Based on \autoref{fig:state_languages}, \autoref{fig:nbBuilds}, \autoref{fig:duration} and \autoref{tab:stars}, we observe a difference of behavior of the projects that have restarted builds.
These projects have a lower build success rate, a much higher number of builds, a much higher build execution time, and are more popular than the projects that do not have restarted builds.
Moreover, those observations are consistent for the top 10 languages, without  exception.
We interpret those differences as the projects with restarted builds are more complex, suggesting perhaps they have more experienced developers that are more likely to use advanced features such as build restart.

\answer{2}{\textbf{Do the projects that restart builds have the same characteristics as other projects that use \travis?} We observe that some languages are more likely to be associated with build restarts. In particular, we observe that JavaScript and TypeScript projects only restart around 1\% of the builds where C++ projects restart 2.82\% of the builds. 
Our hypothesis to explain this difference argues a relationship between programming language and build process complexity.
Simple and faster builds, such as most JavaScript and TypeScript builds, are  restarted less frequently than C++ builds,  for instance.
We also observe that, on average, projects that restart builds differ from projects without restarts in terms of success rate, number of builds, execution time, and popularity.
Our interpretations of these results are that more complex and more experienced projects restart builds more often.}

\subsection{RQ3. Causes of Restarted and Flaky Builds}\label{sec:rq3}

In the third research question, we study the potential reasons that lead developers to restart the builds.
Due to the large scale of our study, it was not possible to manually annotate all restarted builds.
Therefore, we extracted, using \nbRegex\xspace regular expressions, failure reasons from the logs of restarted builds.
We were able to extract failure reasons for {\numprint{\nbFailureReasons}}\xspace builds.
We manually analyzed a sample of the passing builds that were restarted, to further try to surface reasons why developers would restart a passing build.

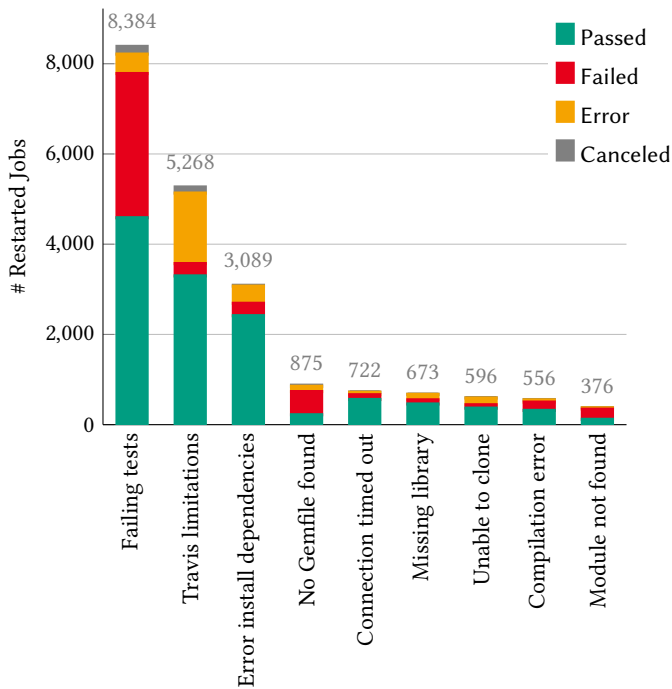
\begin{figure}[t] 
	\centering
	\begin{tikzpicture} 
		\begin{axis}[
			ybar stacked,
			xmin = 0.5,
			xmax = 9.5,
			ymin = 0,
			axis x line* = bottom,
			axis y line* = left,
			ylabel= \# Restarted Jobs,
			width= 0.48\textwidth,
			height = 0.4\textwidth,
			ymajorgrids = true,
			bar width = 4mm,
			xticklabels = \empty,
			extra x ticks = {1,2,3,4,5,6,7,8,9,10},
			extra x tick labels = {Failing tests, Travis limitations, Error install dependencies, No Gemfile found, Connection timed out, Missing library, Unable to clone, Compilation error, Module not found, Connection terminated, port already in use, Unable to push, Checkstyle, Execution, Invalid cmake version, No interactive operation allowed, generic,},
			extra x tick style={
              tick label style={rotate=90}
            },
            nodes near coords={},
            legend style={ 
               row sep=3pt, 
               draw=none, 
               anchor=north,
               cells={anchor=west,font=\sffamily}},
			]
            
            \addplot+[mark=none, gruen_3b, very thick] coordinates {(1, 4586) (2, 3297) (3, 2415) (4, 219) (5, 558) (6, 458) (7, 366) (8, 316) (9, 118) };
            \addplot+[mark=none, rot_9b, very thick] coordinates {(1, 3196) (2, 273) (3, 275) (4, 514) (5, 105) (6, 92) (7, 75) (8, 184) (9, 225) };
            \addplot+[mark=none, orange_7b, very thick] coordinates {(1, 431) (2, 1563) (3, 380) (4, 117) (5, 55) (6, 122) (7, 149) (8, 49) (9, 30) };
            \addplot+[mark=none, gray, very thick,show sum on top] coordinates {(1, 171) (2, 135) (3, 19) (4, 25) (5, 4) (6, 1) (7, 6) (8, 7) (9, 3) };
			
			\addlegendentry{Passed}
            \addlegendentry{Failed}
            \addlegendentry{Error}
            \addlegendentry{Canceled}
		\end{axis} 
	\end{tikzpicture}
	\caption{Reasons of failures of the restarted builds. The state of the restarted build is represented with by the colors, green for passing, red for failure, orange for error, and gray for canceled.}
	\label{fig:failure_reasons}
\end{figure}

\autoref{fig:failure_reasons} presents the ten most frequent types of failure in the execution logs.
Each bar presents the number of jobs that contains the associated type of failure.
Each color represents the state of the build after the builds have been restarted.
For example, it shows that \numprint{8384} builds failed because of test failures, and after the restart \numprint{4586} of those builds are now passing.

We observe that the most frequent cause of build failure is test failure.
This is followed by failures related to the \travis environment, i.e., a build timeout, or no log output for 10 minutes, which has been observed by prior studies as well~\cite{jiang2017causes,labuschagne2017measuring}, 
These results are interesting, because they imply that an important number of restarted builds are restarted for reasons that are related to the source code of the project and not only because of CI environment problems. 
This confirms prior observations~\cite{jiang2017causes}.

Moreover, it is interesting to observe that builds that are restarted in response to test failures have a lower restart success rate, i.e., the original build was failing and it is now passing after the restart, as compared to the other type of failures.
It indicates that developers have more trouble determining which builds to restart when the failures are related to tests.
The other failure types have a much higher success rate of restarts, achieving more than 78\% of successful restart for the \texttt{Error install dependencies} failure type.
This type of failure could, therefore, be automatically restarted with a minimal cost for the CI service.

We also did not expect to see such a large number of passing builds restarted. 
Offhand, there is no obvious reason for a developer to restart a passing build. 
However, 17.17\% of the restarted builds are passing builds that are restarted, and 92.62\% of the time, the passing builds continue to pass after the restart.
We investigated this case, but unfortunately, did not succeed in identifying the reason behind this practice. 
We looked at the type of project, execution time, the difference between the execution times of the original and restarted builds, the \travis configuration, if the restarted passing builds are more related to pull requests.
However, we did not succeed in identifying any specific reason that would explain why developers restart passing builds --- this remains an open question.

\answer{3}{\textbf{Why do developers restart builds?}
We observe that failing test cases are the main reason that developers restart a build.
It is followed by failures that are inherent to the \travis environment, such as execution timeout.
We also observe that the success rate of restarted builds largely varies between the causes of restarts. A build that is restarted after an error during a dependency installation is more likely to succeed than one that is restarted in response to a failing test case.
Further studies should also focus on restarted passing builds, in order to understand developer's goal.}

\subsection{RQ4. Impact of the Restarted Builds}

In this final research question, we explore the impact of restarted builds on the development workflow.

\subsubsection{Developer workflow} 
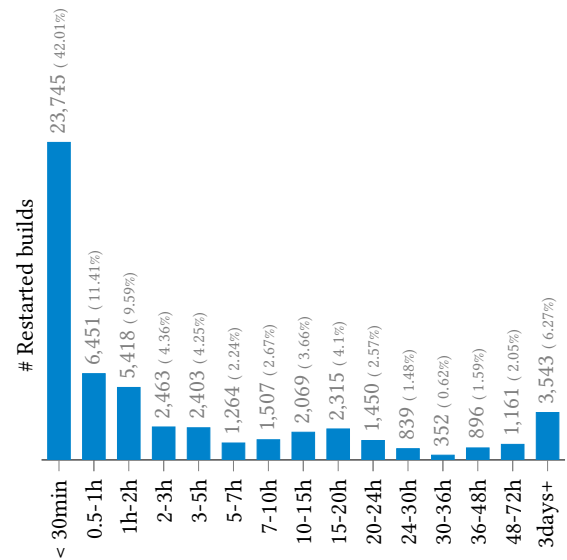
\begin{figure}[t] 
	\centering
	\begin{tikzpicture} 
		\begin{axis}[
			ybar,
			xmin = 0.5,
			xmax = 15.5,
			ymin = 0,
            bar width = 3mm,
			axis x line* = bottom,
            y axis line style= { draw opacity=0 },
			ytick=\empty,
			ylabel= \# Restarted builds,
			width= 0.48\textwidth,
			height = 0.35\textwidth,
			point meta={y*100/\nbRestartedBuilds},
			visualization depends on=rawy \as \myy,
            nodes near coords={\pgfmathprintnumber\myy~{\scriptsize( \pgfmathprintnumber\pgfplotspointmeta\%)}},
			every node near coord/.style={
			    color = gray,
			    rotate=90,
			    anchor=west
			},
			ymajorgrids = true,
			xticklabels = \empty,
            extra x ticks = {1,2,3,4,5,6,7,8,9,10,11,12,13,14,15},
			extra x tick labels = {< 30min, 0.5-1h, 1h-2h, 2-3h, 3-5h, 5-7h, 7-10h, 10-15h, 15-20h, 20-24h, 24-30h, 30-36h, 36-48h, 48-72h, 3days+, },
			extra x tick style={
              tick label style={rotate=90}
            },
			]
			
			\addplot+[mark=none, blau_2b] coordinates {(1, 23745) (2, 6451) (3, 5418) (4, 2463) (5, 2403) (6, 1264) (7, 1507) (8, 2069) (9, 2315) (10, 1450) (11, 839) (12, 352) (13, 896) (14, 1161) (15, 3543) };
		\end{axis} 
	\end{tikzpicture}
	\caption{Amount of time between the original builds and the restarted builds.}
	\label{fig:restarted_time}
\end{figure}

We first look at when, temporally, developers restart builds relative to the original build.
\autoref{fig:restarted_time} shows the distribution of the amount of time separating the beginning of the original build and the beginning of the restarted build.
We see that 42.01\% of the builds are restarted in less than 30min, and 63.01\% of the builds are restarted within 2 hours.
This suggests that a developer saw that a build is failing, analyzed the failed build, and decided to restart it quickly after the end of the original execution.
By implication, this means that developers are either watching the build as it executes, or must stop what they are doing to analyze the build and restart it.
This implies that the developer has to switch between potentially complex tasks, as observed by Czerwinski et al. \cite{10.1145/985692.985715}, and also that they are interrupted, a source of stress, higher frustration, time pressure and effort~\cite{10.1145/1357054.1357072}.
This is particularly an issue when a build is flaky, because these are ultimately unnecessary interruptions that lead to longer wait times and more interruptions.


\subsubsection{Pull Requests}

We now analyze the impact of restarted builds on pull requests. 
A pull request is a code modification (one or more commits) proposed by a developer which is typically reviewed before being merged into the project.
To study the impact of restarted builds, we look at the merging time of the pull requests. 
The merging time is the amount of time between the pull request is opened until it is accepted and merged into the project source code.

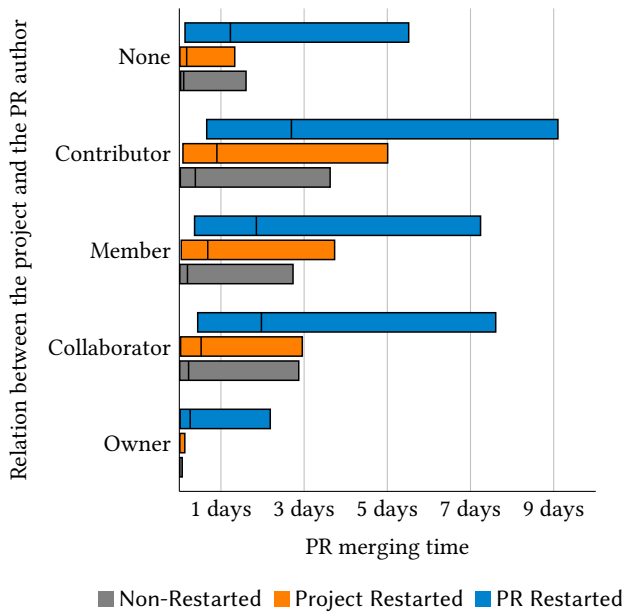
\begin{figure}[t] 
	\centering
\begin{tikzpicture}
\begin{axis}[
    boxplot/draw direction=x,
    xmin=0,
    ymin=0,
    ymax=5,
	width= 0.40\textwidth,
	height = 0.45\textwidth,
    axis x line* = bottom,
    axis y line* = left,
    xlabel= PR merging time,
    ylabel= Relation between the project and the PR author,
    xmajorgrids = true,
    ytick style={draw=none}, 
    xtick style={draw=none}, 
    cycle list={{gray},{orange},{blau_2b}},
    boxplot={
            draw position={1/4 + Floor(\plotnumofactualtype/3) + 1/4*mod(\plotnumofactualtype,3)},
            box extend=0.2
    },
    ytick={0,...,5},
    y tick label as interval,
    yticklabels={Owner, Collaborator, Member, Contributor, None},
    xticklabel style={/pgf/number format/fixed},
    scaled x ticks=false,
    xtick={86400, 259200, 432000, 604800, 777600},
    xticklabel={ 
        \pgfset{fpu}%
        \pgfmathsetmacro\days{floor(\tick/86400)}%
        \pgfmathprintnumber{\days} days%
    },
    legend style={
        row sep=3pt,
        draw=none,
        legend columns=-1,
        at={(0.45,-0.18)},
        anchor=north,
        cells={anchor=west,font=\sffamily}
    },
    legend image code/.code={
        \draw[#1, draw=none] (0cm,-0.1cm) rectangle (0.2cm,0.1cm);
    },
]
\addlegendentry{Non-Restarted\;};
\addlegendentry{Project Restarted\;};
\addlegendentry{PR Restarted\;};

\boxplott{OWNER}{253373.7580024943}{500}{119}{5991.25}{-2874}{120846260}
\boxplott{OWNER}{150004.7092741421}{854}{146}{11377}{-3457}{50554311}
\boxplott{OWNER}{413278.34696406446}{22438}{1683}{188885.5}{5}{64886540}
\boxplott{COLLABORATOR}{471548.7673032232}{19217}{764}{248044.75}{2}{128904887}
\boxplott{COLLABORATOR}{450204.47702921653}{45244}{2607}{255398}{4}{89332207}
\boxplott{COLLABORATOR}{907629.6531440163}{170513}{38136.75}{657166}{6}{96672081}
\boxplott{MEMBER}{582955.188177814}{16872}{951}{236131}{4}{166426864}
\boxplott{MEMBER}{576029.9686449061}{59028}{3977.5}{322543.5}{-1815}{120399936}
\boxplott{MEMBER}{878416.3121793141}{160023}{31622}{625696}{5}{72899680}
\boxplott{CONTRIBUTOR}{620865.5580087104}{33264.5}{2017}{313259.25}{-344}{109277396}
\boxplott{CONTRIBUTOR}{678133.9690322788}{78002}{7484}{432744}{3}{146790492}
\boxplott{CONTRIBUTOR}{956612.6721209288}{232795}{56865}{786371}{-1286}{137047686}
\boxplott{NONE}{486186.0196652453}{9099}{2587}{138538}{-163}{87758286}
\boxplott{NONE}{433773.90417890955}{15204}{1139.25}{115059}{2}{76563661}
\boxplott{NONE}{813130.972972973}{106046.5}{11800.25}{476439.75}{26}{33819815}

\end{axis}
\end{tikzpicture}
\caption{Box plot of the pull request merging time. The $y$ axis presents the relationship between the project and the author of the pull request. The $x$ axis contains the amount of time requires to merge the pull request.
The gray boxes are for the PR in projects that do not have restarted build. The orange boxes are the PRs that are from projects that have at least one restarted build. The blue boxes are the PRs that have restarted build.}
\label{fig:prMergingTime}
\end{figure}

\autoref{fig:prMergingTime} presents box plots of merging time of all pull requests that executed a build during our study. 
The $y$ axis shows the relationship between the author of the pull request and the project: \texttt{None} indicates that there is no relationship between the author and the project, e.g., the author never contributed to it; \texttt{Contributor} means that the author contributed to the project at least once previously; \texttt{Member} means that the author is part of the project organization; \texttt{Collaborator} means that the author has been added as collaborator by the project owner; and finally, \texttt{Owner} means that the author of the PR is the owner of the project itself.
Each relationship type is divided in three: the pull requests from projects that do not have any restarted builds  (gray box), the pull requests from project that have at least one restarted build, but not in the current pull request (orange box), and the pull requests that have at least one restarted build (blue box).
The $x$ axis presents the amount of time requires to merge the pull request.

\autoref{fig:prMergingTime} shows that the merging requires substantially more time when the pull request has a restarted build (blue box). 
This difference cannot be completely explained by a difference of project characteristics , since the difference between the projects that do not have a restarted build and the projects that have a restarted build, respectively the gray and orange boxes, is much smaller.
The median time of the pull request with a restarted build is \textbf{11x} larger than the median time of projects that do not have restarted builds.
It brings the median time of merging time from 4h to 48h.
The merging time is \textbf{3x} larger compared to the projects that have at least a restarted builds, which takes the median merging time from 16h to 48h.

The difference in merging time can be explained by several potential factors. 
The build has to be reexecuted, which requires time to reexecute the build along with the time required for a developer to manually initiate the build restart.  
Moreover, only project members (if they have permission on the specific project), collaborators, and owners have sufficient permission to restart a build.
A contributor therefore needs someone else to restart the build for them.

A second factor is that a restarted build can also start a discussion related to the flaky behavior of the build. 
For example, in pull request \#13477 of keras-team/keras project,\footnote{\url{https://github.com/keras-team/keras/pull/13477\#discussion_r337469914}, visited \today{}}
the author and a member of the project discussed what to do with a flaky test. 
This discussion takes time, and therefore slows down the merging of the pull request.

\answer{4}{\textbf{ What is the impact of flaky builds on development workflow?}
We observe that the restarted builds have an impact on the development workflow.
We firstly show that developers restart builds shortly after the build failed. 
This indicates that developers are waiting for the results or stopping their current task to analyze and restart the builds.
Secondly, we observe that restarted build slow down the merging process of pull requests. We measure a slowdown of 11 times compared to projects that do not have restarted build and three 3 compared to projects that have at least a restarted build.}

\section{Discussion}\label{sec:discussion}

As shown in our results for RQ1, the majority of restarted builds change their status between the original and restarted executions. 
This indicates that developers are relatively successful in identifying which builds would have a different CI result if they are restarted.  
We also show that those restarted builds are associated with a considerably slower merging of pull requests. 
It is, therefore, important to study potential solutions that can be proposed to developers to mitigate this problem.
An obvious mitigation strategy would be to restart certain builds automatically; however, there is a trade-off to balance reliability and cost in terms of execution cost and feedback delay, i.e., sending a notification to the developer.

In RQ3, we observe that the main cause of restart is related to failing tests, followed by causes related to network issues, e.g., connection timeout, and \travis limitations, e.g., execution timeout.
The main cause, flaky tests, is most difficult to identify in terms of whether a build should be restarted.
Indeed, detecting test flakiness requires platform specific techniques that require additional instrumentation and execution \cite{bell2018d}.
This makes it difficult to scale the detection of flaky tests.
At best, any such solution is computationally expensive.
The causes related to network issues and \travis limitations are easier to identify, and present a high restart success rate (see \autoref{fig:failure_reasons}).
Even in those cases, an automatic restart is not a trivial task because the restart should be carefully planned.
It would not make sense, for example, to restart a build that is failing because a website is temporarily out of service; a restart needs to wait for the website to become available. 
We plan in future work to design a new approach to automatically restart builds on \travis and study how developers perceive this type of amortization.


\section{Threats to validity}\label{sec:threats}

\emph{1) Internal: Did we skew the accuracy of our results with how we collected and analyzed information?}
Collecting restarted builds requires constant monitoring of \travis activity, as it cannot be collected after the fact.   
During the study period, we faced several server crashes and a disk crash. 
Those issues forced us to pause data collection for several weeks.
This may introduce non-representativeness in the data.
However, we ran the collection over a large period of time and collected millions of builds, which reduce this risk.

We manually analyzed logs from failing builds to identify regular expressions to identify causes of failure.
There is a risk that we could miss important causes of failure due to the volume of logs that needed to be analyzed.
To mitigate this risk, we provide the complete list of regular expressions in our GitHub repository.
The failing logs and the extracted failure reasons are also provided to allow future researchers to use and verify our work.

\noindent\emph{2) External: Do our results generalize? }
A potential threat is that the study period is non-representative due to abnormal activity on \travis.
We reduce this risk by collecting data during a large period of time, from \fromDate\xspace to \toDate. 
During that period, we monitored the \travis and GitHub announcements to ensure that nothing abnormal was happening.

To enable our results to generalize as much as possible, we selected a large and diverse set or projects in various languages, across a wide range of projects on GitHub.  However, all of our projects are open source, so we cannot make claims about how our results might generalize to proprietary projects.

\noindent\emph{3) Replicability: Can others replicate our results?}
To support others in replicating our results, we make our data and infrastructure code available for other researchers and users.
The source code of \travislistener is available on GitHub with the DOI:
\href{https://github.com/tdurieux/Travis-Listener}{10.5281/zenodo.3709181}.
The data is available on Zenodo with the following DOI: \href{https://doi.org/10.5281/zenodo.3601137}{10.5281/zenodo.3601137}.

\section{Related work} \label{sec:related-works}

This section presents the related works of this contribution.
We focused on two research fields: the works related to \travis and the works related to flaky builds.

\subsection{Related work studying \travis}
Beller et al. \cite{Beller:2017:OMT:3104188.3104232} exploits TravisTorrent to study the build behavior of the projects that use \travis.
Hilton et al. \cite{hilton2016usage} study the use of continuous integration in open-source projects.
They show that continuous integration has a positive impact on the projects, and it is used in 70\% of the most popular projects on GitHub.
Zhao et al. \cite{zhao2017impact} study the impact of \travis on development practices. Their main finding is that GitHub pull requests are more frequently closed after the integration of \travis.
Widder et al. \cite{widder2018m} present a study analyzing the reasons projects leave \travis. 
They observed that this phenomenon is related to build duration and  repository language.
They showed that C\# repositories were more likely to quit \travis because \travis (at the time) did not support Windows virtual machines (support has since been added). 
On the contrary, repositories that have long build are more likely to continue to use \travis.
Durieux et al. \cite{durieux2019analysis} present a study on 35 million jobs of \travis.
They analyze how developers are using \travis and for which purposes.
They find that \travis is still mainly used for building and testing applications. 
Compared to these prior studies, our study targets different aspects of the CI process: We focus on restarted builds and their impact on the development workflow.

Rausch et al. \cite{rausch2017empirical} present a study on 14 open-source projects that use GitHub and \travis. 
They analyzed the build failure and identified 14 different error categories.
They presented several seven observations, such as ``authors that commit less frequently tend to cause fewer build failures'', or ``Build failures mostly occur consecutively''.
The difference between this study of build failure and our study is that this related work focuses on the context of the build failure, where we are interested in the cause of failure, e.g., which errors produce build failures, as well as build flakiness.

\subsection{Related work studying flaky tests}

Luo et al. \cite{luo2014empirical} presented the first empirical analysis of flaky tests. 
They manually analyzed 201 commits from 51 Apache projects that fix flaky tests.
Based on this analysis, they produced a taxonomy of the most common root causes of the flaky tests, as well as identify strategies to identify and fix certain types of flakiness.
Compared to this work, we focus on flaky builds, studying what causes  build flakiness and flakiness impact on development workflow. 
In future work, we plan to analyze the flaky tests that we mined and verify that Lui et al.'s taxonomy can be applied to them.

Eck et al. \cite{eck2019understanding} study the developer's perspective on the flaky tests. 
They build a dataset of 200 flaky tests and then ask the original developers that fixed the problem to analyze the nature, fixing effort and the origin of the flakiness. 
Developers indicated that flaky tests are rather frequent and a non-negligible problem.
They only considered that reproducing the context leading to the test failures and understanding the nature of the flakiness are the most important and most challenging needs.
In our study, we did not contact the developers to understand their perspective on the flaky tests. 
However, we provide a dataset and framework that could help the community to further investigate flaky tests.

Bell et al. \cite{bell2018d} proposed DeFlaker, an automated technique that identifies flaky tests by running a mix of static and dynamic analyses.
Lam et al. \cite{lam2019idflakies} presented iDFlakies, a tool that identifies tests that are flaky due to order execution.
Compared to those two contributions, our contribution does not aim to detect automatically flaky tests. 
However, a new benchmark of flaky tests could use our approach of restarted build detection to identify projects that have potential flaky tests.  

Micco \cite{micco2017state} presents the current state of continuous integration at Google.
He indicates that 84\% of transitions from passing to failing tests at Google are from "flaky" tests. 
During our study, we did not observe the same behavior in the open-source project. 
This probably due does a different nature of the testing that is made at Google compared to testing in open-source projects.

\section{Conclusions} \label{sec:conclusions} 

In this study, we analyzed the restarted builds from \travis, one of the most popular build systems.
We collected \numprint{\nbRestartedBuilds}\xspace restarted builds from \numprint{\nbRestartedRepositories}\xspace projects. 
We show that \percentSuccessfullRestart\% of the restarted builds are passing. 
We identify older and more complex projects are more prompted to restart a build.
Interestingly, the different programming languages are not subject to the same level of build restart. 
JavaScript and TypeScript are the languages with the lower restart rate, and C++ is the language with the most frequent request.

We identify that test failures and tests in error are the most common causes of build restarts, followed by the causes that are related to the \travis environment, i.e., execution timeout.
We also identify that the different failure reasons have a different restart success rate.

Moreover, we observe that a restarted build entails an impact on the development workflow.
Developers need to wait for the outcome of the builds and therefore introduce a serious delay when the build needs to be restarted several times.
Moreover, we observe that restarted builds introduce a serious delay in pull request merging. Indeed, on average, the pull requests that have a restarted build are 11 times slower than the pull requests from projects that do not have restarted build.

Finally, we discuss the possibility of restarting fully automatically some builds that are failing due to reasons that have a high restart success rate. 
This automatization would save developers' save time and effort, hence reducing the delay of the contribution of an external contributor.

\begin{acks}
This material is based upon work supported by Funda\c{c}\~ao para a Ci\^encia e Tecnologia (FCT), with ref. PTDC/CCI-COM/29300/2017 and UIDB/50021/2020, and the National Science Foundation (CCF-1750116). Thomas Durieux was partially supported by FCT/Portugal through the CMU-Portugal Program.
\end{acks}

\bibliography{references}
\balance

\end{document}